\documentclass[sigplan,10pt]{acmart}
\renewcommand\footnotetextcopyrightpermission[1]{}
\setcopyright{acmlicensed}
\copyrightyear{2024}
\acmYear{2024}
\usepackage{xspace}
\usepackage{booktabs} % for professional tables
\usepackage{tikz}
\usetikzlibrary{positioning,shapes.geometric,arrows,fit,shapes.symbols,svg.path,tikzmark,calc}
\usepackage{textcomp}
\usepackage{listings}
\usepackage{subcaption}
\pgfdeclarelayer{background}
\pgfdeclarelayer{foreground}
\pgfsetlayers{background,main,foreground}
\usepackage{siunitx}
\usepackage{adjustbox}
\usepackage{array}
\usepackage{multirow}
\usepackage{cleveref}
\usepackage{paralist}
\usepackage{array}

\usepackage{amsthm}
\newtheorem{definition}{Definition}
\usepackage{siunitx}

\usepackage{algorithm}
\usepackage{algpseudocode}

\let\labelindent\relax %https://tex.stackexchange.com/questions/170772/command-labelindent-already-defined

\usepackage[inline]{enumitem} % for inline lists, e.g. \begin{enumerate*}[label=\roman*)]  See https://tex.stackexchange.com/questions/146306/how-to-make-horizontal-lists

\newcommand{\algname}{PSL\xspace}
\newcommand{\dc}{storage server\xspace}
\newcommand{\dcs}{storage servers\xspace}

\newcommand{\seq}{PSL-DB\xspace}

\newcommand{\Seqs}{PSL-DBs\xspace}
\newcommand{\Seq}{PSL-DB\xspace}

\begin{document}

\title{
Lightweight, Secure and Stateful Serverless Computing with PSL
}

\author{Alexander Thomas}
\email{alexthomas@berkeley.edu}
\affiliation{%
    \institution{UC Berkeley}
    \country{USA}
}

\author{Shubham Mishra}
\email{shubham\_mishra@berkeley.edu}
\affiliation{%
    \institution{UC Berkeley}
    \country{USA}
}
\author{Kaiyuan Chen}
\email{kych@berkeley.edu}
\affiliation{%
    \institution{UC Berkeley}
    \country{USA}
}
\author{John Kubiatowicz}
\email{kubitron@berkeley.edu}
\affiliation{%
    \institution{UC Berkeley}
    \country{USA}
}

\begin{abstract}
    
% Heterogeneous infrastructures, such as multi-cloud, with varying security policies necessitate
% Trusted Execution Environment (TEE), which provides confidentiality and integrity and strong isolation from the untrusted kernel or infrastructure. 
% Due to the complexity of its programming models and underlying infrastructure management, Function-as-a-Serivce (FaaS) becomes a viable option to simplify the development and deployment of security-critical applications to TEE. 
% Providing an interface to FaaS for confidentially storing and exchanging states reduces the risk of inadvertent data leakage due to developer error.
% Adapting conventional TEE execution frameworks, such as Library OS-based containers or interpreted language engines, leads to inefficiencies and even potential security breaches.

% Trusted Execution Environment (TEE), such as Intel SGX and AMD SEV, provides confidentiality, integrity and strong isolation from the untrusted infrastructure providers. However, its complexity makes Function-as-a-Serivce (FaaS) to be a favorable option for security-critical application development and deployment.
We present \algname, a lightweight, secure and stateful Function-as-a-Serivce (FaaS) framework for Trusted Execution Environments (TEEs). 
The framework provides rich programming language support on heterogeneous TEE hardware for statically compiled binaries and/or WebAssembly (WASM) bytecodes, with a familiar Key-Value Store (KVS) interface to secure, performant, network embedded storage.
It achieves near-native execution speeds by utilizing the dynamic memory mapping capabilities of Intel SGX2 to create an in-enclave WASM runtime with Just-In-Time (JIT) compilation. 
\algname is designed to efficiently operate within a asynchronous environment with a distributed tamper-proof confidential storage system, assuming minority failures.
The system exchanges eventually consistent state updates across nodes while utilizing release-consistent locking mechanisms to enhance transactional capabilities.
The execution of \algname is up to 3.7x faster than the state-of-the-art SGX WASM runtime. 
\algname reaches 95k ops/s with YCSB 100\% read workload and 89k ops/s with 50\% read/write workload. 
We demonstrate the scalability and adaptivity of \algname through a case study of secure and distributed training of deep neural network.

\end{abstract}

\maketitle
\pagestyle{plain}

\vspace*{-0.1in}
\section{Introduction}

% 

% if people have their own data, if they wnat to do it securely, here is how they might launch it. launch a training, result put back, doesn't leak  

% \begin{figure}
%     \centering
%     \includegraphics[width=0.8\linewidth]{figures/Ring Overview.pdf}
%     \caption{Paranoid Stateful Lambdas (PSLs) exploit pervasive and
%       abundant computational resources from the edge and cloud, thus
%       enabling a universal plane of secure and stateful in-enclave
%       execution. The Secure Concurrency Layer (SCL), provides coherent
%       and eventually-consistent semantics over data writes between PSLs,
%       while ensuring durability using network-embedded DataCapsules. \eric{placeholder, need to make the figure more relevant}}
%     \label{fig:overview}
% \end{figure}

\begin{figure}
    
    \centering
    \includegraphics[width=0.9\linewidth]{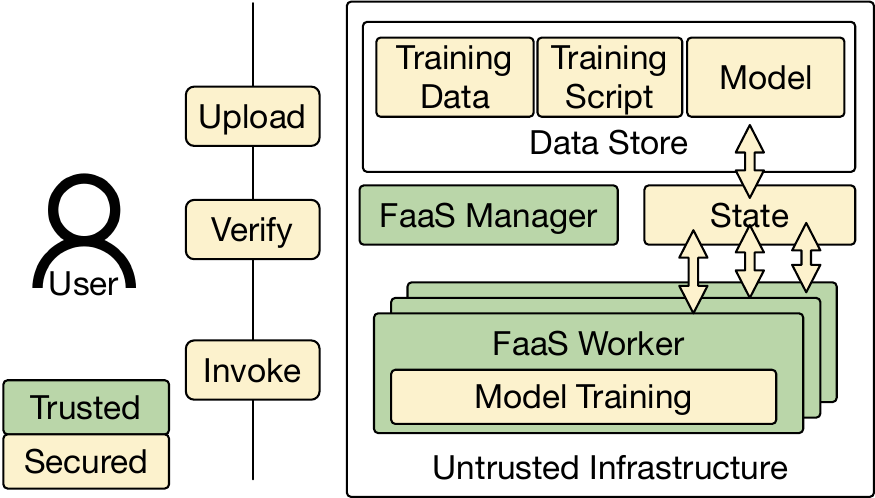}

    \caption{\textbf{Use Case of \algname in Distributed Learning Training on Confidential Data}. In \algname, we consider privacy-preserving distributed deep learning training on untrusted infrastructure. Contrary to the conventional secure enclave programming interface that forces users to protect their data with hand-crafted remote encryption schemes, \algname provides Function-as-a-Service (Faas) with simple and easy-to-use interfaces, in which users exploit standardized mechanisms to \emph{store} the function and parameters to servers, \emph{verify} that an expected function will be executed with confidentiality and integrity, and \emph{invoke} the function.  }
    \label{fig:intro:usecase}
    \vspace*{-0.2in}
\end{figure}

% Heterogeneous infrastructures, such as multi-cloud, with varying security policies necessitate
% \eric{TEE is good}
Trusted Execution Environments (TEEs) are becoming a popular way to deploy security-critical applications in untrusted environments. TEEs, such as Intel Software Guard Extensions (SGX) and AMD Secure Encrypted Virtualization (SEV), execute privacy-preserving applications in \emph{secure enclaves} with confidentiality, integrity, and strong isolation from the untrusted kernel or infrastructure. 
% \eric{working with tee is hard, so faas is more suitable} 
Unfortunately, developing and managing applications on TEEs can be error-prone and complex. Inadvertent programming errors can result in long development cycles and private data leakage.  

In contrast, serverless computing, with the Function-as-a-Service (FaaS) model, is an emerging computing paradigm that streamlines application development and deployment without the complexity of building and maintaining the infrastructure.
With FaaS, one can decompose data-intensive applications into many functions and exploit massive parallelism and scalability of cloud infrastructure. 

In this paper, we show how FaaS becomes an effective option for TEE-based applications by offloading the burden of managing underlying cryptographic hardware infrastructure and state management to the framework. %thus simplifying the development process for security-critical applications and minimizing the risk of human error.
Our secure FaaS framework scales easily across different computing infrastructures, such as multi-cloud and edge-cloud environments, with varying security policies. 
 
% \eric{statefulness is also more important to tee's faas}
\textit{Statefulness}, a property provided by some of the FaaS frameworks, such as Faasm~\cite{shillaker2020faasm} and Cloudburst~\cite{sreekanti2020cloudburst}, allows applications to get access to a Key-Value Store (KVS) to exchange the states with other FaaS workers. Instead of relying on some external database service, stateful FaaS enables state sharing natively, thereby increasing system availability and lowering the latency of propagating state updates.  Statefulness is useful for data-intensive applications, such as video processing~\cite{zhang2019video}, 
machine learning~\cite{carreira2018case}, and 
robotics~\cite{ichnowski2019mpt, ichnowski2022fogros}. %It enables flexible sharing of , such as chaining.  
However, the \emph{way} in which statefulness is integrated into a FaaS environment 
%Providing an interface to FaaS for confidentially storing and exchanging states is 
is a crucial consideration for in-enclave applications to reduce the risk of inadvertent data leakage. % due to developer error.

By way of example, Figure \ref{fig:intro:usecase} shows a distributed learning algorithm that trains on private data. A stateful and secure FaaS framework can streamline the development and deployment of this application
%by providing a simple stateful FaaS interface to the user. It can shield 
by shielding the user from having to manage the enclave environments of multiple workers, verify that workers are running correct code, protect and secure training data through encryption and signatures, and schedule communication for updates to model data. 
% \shubham{Some motivating example application case-study here? Basically need to say stateful FaaS "in-enclave" is a useful idea.}
% A secure and stateful in faas should also 
%, which removes inf. from the trust model

% We consider heterogeneous cloud environment ... \eric{one sentence for problem formulation, motivate bft here?} 
Challenges in stateful FaaS have been greatly discussed and mostly addressed in out-of-enclave setting, such as ~\cite{shillaker2020faasm, sreekanti2020cloudburst}, and challenges in providing a usable execution framework have been proposed such as Graphene~\cite{tsai2017graphene}, Occlum~\cite{shen2020occlum}. However, limited systems have been proposed in in-enclave stateful FaaS due to the following challenges: % \eric{any justification on this vs occlum?}

% \begin{compactenum}[\bfseries 1.]
%\begin{enumerate}[wide, labelwidth=!, labelindent=0pt]

\paragraph{Execution Model:} Heterogeneous TEE hardware involves varied programming capabilities, leading to a trade-off among \emph{language support}, \emph{portability}, and the capability to \emph{isolate the application from the runtime}. 

For instance, AMD SEV provides secure enclaves as confidential VMs, thus programming akin to standard cloud virtual machines. In contrast, Intel SGX\footnote{SGX is available for all Intel processors before 2022, and currently available for all Intel Xeon servers.  
This work assumes its latest generation (SGXv2). } , one of the dominating secure enclaves available on Intel processors, 
requires that predefined functions calls from untrusted operating system to enclave and from enclave to untrusted operating system. While this reduces the Trusted Computing Base (TCB) by only including required components, it necessitates specialized designs and modifications to both the program and its associated libraries. 

In the basic SGX model, the binary is required to be statically linked to ensure the loaded code pages cannot be modified, leading to large binaries and limited modularity and portability.  The basic model also does not provide confidentiality of the executable -- a significant limitation in today's untrusted environments.

Providing an efficient execution model that unifies heterogeneous TEE hardware is important. Existing work uses interpretation: for example, S-FaaS~\cite{alder2019sfaas} only supports interpreted runtimes like Javascript and WebAssembly Micro Runtime (WAMR) or use ahead of time (AOT) compilation that requires compiling to the correct target CPU and special care needed to work inside secure TEEs.

% and consumes significant compute resources. 

 % This is advantegeous that one can only include specified TCB to their code. However, programming on SGX requires specialized design on the program itself and its associated libraries. Having a performant framework for secure enclaves, especially with both SEV and SGX, is important.

\paragraph{State Security:}
Since TEEs do not guarantee secure or trusted storage, state out of the enclaves needs to be confidential and tamper-proof.
This requires encryption and signatures on the critical path of propagating state updates.
Naively applying conventional encryption and signature schemes leads to an inefficient protocol with compute-bound bottlenecks.
Any system using such cryptographic scheme should also have a dedicated key management infrastructure.
% Any system that supports such a setup not only has to optimize the use of encryption and signature schemes in the state-update protocol but also needs to have a subsystem dedicated to handling key management. \eric{@Shubham, I would write it as: Directly applies conventional storage and state consistency protocols lacks (do not use passive sentences in a paper) dedicated cryptographic key management (key could mean kvs or crypto's key), thus unable to provide state confidentiality and integrity.}
% E.g. anna's eventual consistency with CRDT to reduce the overhead of sync. However, in enclave is different due to the trust model. \eric{motivate bft}

\paragraph{Consistency:}
Existing TEE-based consistent state management generally use consensus protocols for total order over all operations being executed in the system.
We argue that this is not the perfect fit for FaaS. Firstly,  not all applications need a total order. Using a leader-based consensus protocol like Raft \cite{ongaro2014search} (as used in CCF \cite{ccf_new_vldb}), leads to lower overall throughput as only the leader node is allowed to propose writes. 
% \eric{put this sentence to the front, total order, such as ccf However, for faas, we don't need ...}
However, the strong total order and consensus might limit the scalability of FaaS workers. 
% FaaS application without the need for this strong total order guarantee might want to scale out by adding more FaaS workers, which is not possible with consensus.
Secondly, the churn in a FaaS environment is supposed to be much higher than a distributed database setting. Consensus protocols generally have a reconfiguration phase which causes a node joining the system to have some added startup latency.
This latency is critical for FaaS applications that are generally short-lived.
% Thus, we believe, it is imperative for a general-purpose stateful FaaS framework to use a more relaxed eventually consistent model.

% \end{enumerate}

To resolve these challenges, we observe the need for a TEE-based stateful FaaS framework that:
(1) supports multiple languages and TEEs with low overhead, while also protecting the confidentiality of the application itself;
(2) isolates runtime from the application workers;
(3) supports efficient confidential state sharing with options for locking, but only when necessary;
(4) guarantees durable eventually consistent state updates. 

% In this work, we present an end-to-end solution for building a secure stateful Function-as-a-Service (FaaS) platform using Trusted Execution Environments. We have three contributions: 
% Second, Thirdly, 

% We achieve secure state management with BFT environment. We use a distributed tamper-proof confidential storage system to coordinate eventually consistent state updates among FaaS workers with release consistent locking for transactional capabilities.

% implementation 

% results 
% We evaluate \algname with \eval{}. We show a case study by porting a distributed deep neural network training to \algname. We show \algname scales with increasing number of workers. 

% contribution 
To this end, we present \algname, a secure, lightweight and stateful FaaS framework. \algname has the following properties, summarizing our contributions: 
\begin{itemize}
%[wide, labelwidth=!, labelindent=0pt]
    % \item \textbf{Lightweight Enclave Execution with JIT Compilation}
    % \item \textbf{Secure and Efficient State Sharing}
    % \item \textbf{Evaluation Data of \algname and A Case Study on Distributed LLM Training}

    \item \textbf{\algname supports a portable binary with native support for multiple programming languages and heterogeneous enclave hardware}. It uses create an in-enclave WebAssembly (WASM) runtime to provide multi-language support at near-native speed. It provides support for popular secure hardware backends (Intel SGXv2 and AMD SEV) which can be extended to other secure hardware given a list of supported primitives. 
    \item \textbf{\algname enables Just-In-Time compilation that out-performs state-of-the-art SGX WASM runtimes.} Unlike traditional use of SGX, our system also permits statically compiled binaries to be encrypted for privacy and unpacked only within the protections of an enclave.  We use the dynamic memory mapping capabilities of Intel SGX2 with just-in-time compilation. 
    % \item \textbf{\algname supports scalable and efficient in-enclave state management} \eval{single machine latency, throughput }
    \item \textbf{\algname enables secure state persistence in untrusted storage.} Data outside the enclaves is always encrypted, hashed, and signed, and durably stored in a majority quorum of \dcs. 
    % \eval{consistency metrics: multiple machine latency, throughput, violation rate}
    % passive replication in bft env. 
    \item \textbf{\algname supports scalable, efficient, and eventually-consistent state updates with release-consistent locking.}
    We provide a state update protocol that works in an asynchronous network environment and performs well even though updates are signed, hashed, and encrypted for secure storage in the network. With 8 FaaS workers, it provides a throughput of 89k ops/s with a 50-50 read-write YCSB workload with all cryptographic protocols enabled. 
    % fault tolerant state persistence; passive replication \eval{}
    %  \eric{not sure if we can evaluate this one; faasm also has this contribution as "flexible host interface" } 
    % \eric{We achieve efficient confidential state management. State updates propagate in an eventually consistent way in an asynchronous Byzantine network environment.
\end{itemize}

\section{Background and Motivation}

\begin{figure*}
    \centering
    \includegraphics[width=0.8\linewidth]{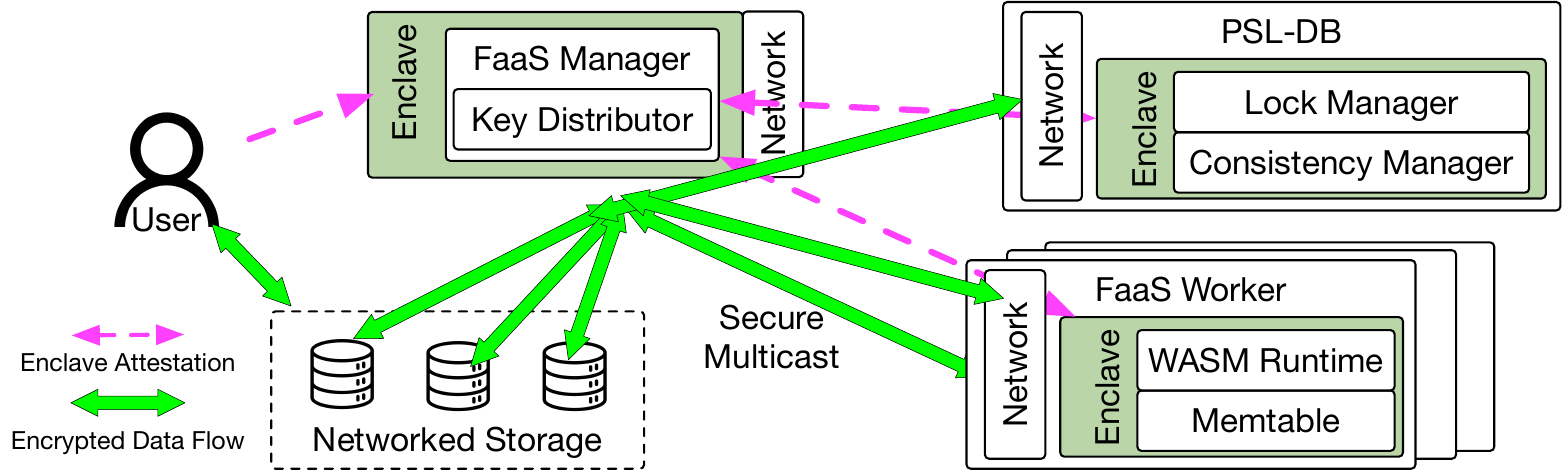}
    \vspace{-0.15cm}
    \caption{\textbf{The architecture of PSL:} The FaaS Manager launches and attests all worker enclaves.
    Enclave workers cache key-value pairs in their local Memtable, and share state in an eventually consistent way using a Secure Concurrency Layer (SCL) that takes advantage of a secure-multicast primitive within the network. The \Seq tracks the most recent version of the state of each key and pushes updates to a set of replicated networked \dcs that provide state persistence.
    }
    \label{fig:overview}
    \vspace*{-0.1in}

\end{figure*}

We introduce the background of secure enclave execution and discuss the limitations of existing approaches in current secure and stateful FaaS frameworks. 
%\eric{Note: the four section names are placeholders. We need to come up with more accurate sectioning}

% \begin{figure}
%     \centering
%     \includegraphics[width=\linewidth]{figures/feature_comparison.png}
%     \caption{\textbf{Feature Comparison of different enclave execution models} \eric{language support}  \eval{numbers on some of the columns: execution speed , loading speed, fault binary size (wasm compared to alternatives)} \eric{goal is to show WASM is a logical choice for enclaves} \eric{TCB}}
%     \label{fig:motivation:comparison}
% \end{figure}

% \begin{figure}
%     \centering
%     \includegraphics[width=\linewidth]{figures/binary_size.png}
%     \caption{\textbf{Distribution size comparison of different enclave execution models} \eric{potentially merge with feature comparison and make it to be double column}}
%     \label{fig:motivation:binary_size}
% \end{figure}

\subsection{Execution Models in TEE}

Trusted Execution Environments (TEEs) create \emph{enclaves} which are secure, isolated environments protected from the privileged host OS, hypervisor, and any hardware devices connected to the host.  There are at least two different classes of TEEs in wide use.

The first is the \emph{confidential VMs} as provided by AMD SEV and the upcoming Intel TDX.  These systems provide a programming model akin to standard cloud virtual machines with extra protection. 
Confidential VMs typically involve a large trusted computing base (TCB), where the user needs to trust all the binaries and drivers in the VM image provided by the infrastructure provider.  The challenge in this environment is to \emph{reduce} the TCB as much as possible.

The second is exemplified by Intel SGX, which provides a more restricted programming model that provides a secure execution container more akin to a process and that requires predefined function calls from the untrusted operating system to enclave (ecalls) and from enclave to untrusted operating system (ocalls). While this structure reduces the TCB by only including required components, it necessitates specialized designs and modifications to both the program and its associated libraries.  

Providing an execution model for confidential VMs is relatively straightforward, since any VM image that works on typical cloud machines should work with minimal changes and performance degradation in a Confidential VM domain.  As a result, we focus on providing an execution model that works for Intel SGX. We then extend our SGX-driven execution model to Confidential VM environments by providing a restricted unikernel-based runtime with minimal functionality to mirror the needs of our SGX environment.

Existing execution models in Intel SGX can be typically categorized
as following: 
\begin{enumerate}
    \item \emph{Static linking}:  Conventional SGX usage statically links all the dependencies within a single binary. SGX requires loading all code pages at startup time for code integrity, which incautious design may end up loading with gigabytes of the binary.  Conventional SGX usage also prevents binaries from being encrypted, thereby exposing algorithms to external analysis.   
    \item \emph{Interpreter}: Some frameworks, such as S-FaaS, statically link a language interpreter such as Javascript. This approach provides limited language support, and cannot support more complex interpreted languages such as python. 
    % \item \emph{LibOS-loader}: existing work implements minimal Library OS that typically proxies the system calls to the host system out of the secure enclave. In LibOS, one can use as Linux with only one process, 
    \item \emph{Library OS (LibOS)}: Existing work implements minimal Library OS that typically proxies the system calls to the host system out of the secure enclave. With Library OS, one can use as Linux with only one process. Although existing container orchestration frameworks (\emph{i.e.}, Kubernetes), can be utilized here, 
    %One can use existing container orchestration frameworks, such as %Kubernetes, to manage. However, 
    %whenever one wants to switch to another application,
    the container must be reloaded to switch to another application. 

    \item \emph{Sandboxed Runtime}: Using a sandboxed runtime has become a more popular option, given the rise of WebAssembly. Recent systems compile WebAssembly MicroRuntime (WAMR) to SGX enclaves to dynamically interpret and execute WebAssembly. 
    % \item \emph{WebAssembly MicroRuntime (WAMR)} Dynamically interprets and executes WebAssembly execution 
    % \item \emph{WebAssembly AoT}: Large binary
    % \item \emph{WebAssembly Virtual Machine (WAVM)} Executes WASM binary with almost native speed with Just-In-Time compilation
\end{enumerate}
% Among all the available options, we make the following observation:

% \textbf{Insight I: } WAVM support is the execution model for FaaS in secure enclaves that best represents the trade-offs. It supports portable binaries compiled from multiple languages for different secure enclave types. It also isolates runtime from the guest applications. 

% Executing WAVM in AMD SEV is trivial. However, to the best of our knowledge, there is not a WAVM execution framework for Intel SGX due to the following challenges: 
% \begin{itemize}
%     \item WASI support
%     \item JIT for performance
% \end{itemize}
% \eric{@Alex, please help to fill this in, this relates which design you would like to focus on}

\subsection{In-Enclave Runtime}
The Function-as-a-Service (FaaS) model in cloud computing enables users to access the cloud infrastructure without configuring it.  Consequently, we believe that FaaS is an attractive option for users wanting the security of TEEs without the hardship of configuration. Conventional use of SGX by separating trusted and untrusted components of code and statically linking with libraries that are compatible with a target TEE is at odds with FaaS; thus, we automatically rule out static linking of this type. Although LibOS can allow for running unmodified applications, it does not provide an execution environment. While an interpreter is generally an attractive option for FaaS, it provides lackluster performance compared to running native code.
We focus on WebAssembly with its rich language support and maturity of different runtimes that advertise near to native speeds.

% Furthermore, a runtime enables users more flexibility in the languages that they support. Using a runtime can also function as a secure loader that can decrypt previously encrypted binaries. This property is useful for when user code must be confidential (i.e. proprietary software). Lastly, 

% We push that FaaS is more than just a paradigm to take the infrastructure away from the execution, but also to ensure users can execute code without not knowing  underlying the infrastructure. We focus on WebAssembly with its rich language support and maturity of different runtimes that advertise near to native speeds. 

% Thus, using lightweight sandboxed runtime is a logical choice to be used as the execution model for FaaS.

% In order to support FaaS model with high utilization, an efficient runtime should statically load as minimum as possible but still guarantee confidentiality, integrity, and isolation of the guest application. 
% Recent studies ~\cite{} show running WebAssembly in a sandboxed runtime is more compelling in FaaS environments than containers, due to the lightweightness and fast loading time. This property is desirable for Intel SGX-based secure enclaves. 

% Intel SGX requires static loading of all binaries at initialization time to generate the hash of the loaded pages in \emph{measurement report}. The measurement report is used to \emph{attest} to the user that the expected and untampered code is loaded and executed in the secure enclave.   

\subsection{A Case for a FaaS JIT Runtime}
\label{subsec:case_for_jit}
% \eric{this section covers WASM JIT can be useful in enclave context with fault isolation and program distribution, but challenging}  
% WASM has multiple runtimes and compile modes. \eric{background on JIT/AOT and their tradeoff} \eric{cite Twine on IWASM}

% We run the aforementioned models an iterative FaaS task that \eval{Alex and I are discussing a reloading test for these models, showing reloading makes a difference due to the runtime. Not sure if we have enough time to do it}. We find execution models have significantly different utilization rates ,, executing in Intel SGX compared to executing out of the enclave.\eric{maybe put attestation background before the execution model} 

WebAssembly supports three execution modes: (1) runtime interpretation that directly interprets bytecode without first compiling to native machine code, (2) Ahead-Of-Time (AOT) that compiles WASM bytecode into native machine code, and (3) Just-In-Time (JIT) that compiles WASM bytecode to native code at runtime. 
Runtime interpretation can overcome the static linking requirement posed by SGX, because interpretation avoids the overhead associated with compiling code and the code is executed within the interpreter itself; however, the cost of decoding instructions on the fly is a significant overhead. 
AOT compiles WASM to native code that is ready to execute as soon as it is loaded with a similar mechanism as Intel SGX's static loading but requires knowing the right flags to configure and the correct libraries to link that are compatible within an enclave.

% It takes longer loses the flexibility of dynamically loading the binaries and needs to reload the enclave to restart. 
In contrast, JIT allows users to compile a single binary that can be run on any JIT-enabled runtime. The JIT runtime can automatically link the correct dependencies and compile the code to become compatible with its secure hardware backend. Furthermore, JIT's compilation process can utilize dynamic information, potentially optimizing the code based on runtime conditions and the specific hardware it is running on.
There is also a security advantage when using a JIT runtime, as using JIT ensures that malicious or unknowing clients with a faulty compiler cannot break the WebAssembly sandbox. Compiling things inside a TEE adds further assurance that the sandbox doesn't become compromised. There has been work done on verifying the sandboxing of WASM binaries \cite{johnson2021}, but it currently is for nightly builds and is not suitable for a low latency FaaS environment.
With all that said, JIT doesn't come for free, as the drawback to using JIT is poor cold start-up time, which is critical in a FaaS environment. We attempt to optimize start-up time latencies to make it more practical in a FaaS environment.  

% Traditionally, what people do is to have an in-enclave program to bootload the actual program. Because fundamentally it does not allow hierarchical trust in enclave. They do it so that user can attest the code. This has the following problems (1) no protection (2) same level of trust 
% ? why aws allows coding instead of binaries 
% Through attestation, a multi-PSL application can be served by an untrusted, third-party service provider. 

% Using JIT ensures that malicious clients or unknowing clients with a buggy compiler cannot break the sandbox as WASM bytecode is translated in-enclave. There is current work that attempts to verify the sandboxing of WASM binaries, but it isn’t suitable in a FaaS environment and mostly made for nightly builds. 

\subsection{State Management}

% \subsection{Eventually Consistent Confidential Shared State}
% \shubham{This needs changes, TODO for me.}
% \eric{this section covers existing stateful protocols are not enclave aware and lead to bad performance}
% FaaS enables parallel execution of a scalable number of workers to compute simultaneously. In some applications, such as big data analytics, X and X, need some level of statefulness to store the intermediate states to the ephemeral storage.  Typically, people choose KVS as the interface, instead of file system, because the interface is easy to use, and the lambda's file system gets preempted for very short period of time.

% There has been a lot of work on stateful faas, such as X, X and X. Figure X shows the performance if directly porting those state storage to secure enclaves. This leads to performance degradation. \eval{a lineplot of moving redis(faasm) to enclave, and the curve quickly flattened} \eric{refer to Speicher's paper}
% This is because enclave has 128MB EPC, which page in and page out takes significant overhead. Paging is implicit so porting applications naively lead to serious performance degration. 

% \eric{this section covers state consistency }
% applications such as X, X and X that need to share the intermediate states with others
% \eric{what do we want to evaluate? }
% Working inside of the enclave is challenging. 
% Most of them focus on the setting out of the enclave, which security guarantee is limited.

While secure enclaves protect the data during computation, special care must be taken to protect the confidentiality and integrity of data at rest and in transit.
Scaling up with multiple FaaS eventually requires sharing of state among the workers.
This raises interesting consistency questions and query performance issues as blocks of data as a whole must be exchanged or stored encrypted and, therefore, can't be queried at a finer granularity.

Management of state in FaaS generally takes one of three forms:
(1) Stateless: The workers use an external database service to store and retrieve state.
(2) Centralized: The workers use an in-cluster database to coordinate state updates among themselves.
(3) Decentralized and eventually consistent: The workers periodically exchange their local cache with each other and merge their updates using a Conflict-free Replicated Datatype (CRDT).
For use in enclave-based FaaS, the decentralized approach is a better fit due to its higher resilience to failures and attacks.

Another line of work runs State Machine Replication (SMR) within enclave-based systems.
While this is a similar approach to the decentralized case above, it provides a stronger total order guarantee, which may not be necessary for many applications.

% Maybe remove this if it consumes space.

\section{Overview}
\label{sec:overview}

We assume a 
heterogeneous compute environment with varied security policies and hardware configurations.
For example, components of \algname could be spread over multiple geo-distributed cloud regions, even going across multiple cloud providers.
Another example of a target heterogeneous environment is FaaS workers running in an edge environment where the data is persistently stored in the cloud.

\subsection{Threat and Network Model}
\algname adopts the typical cloud attackers who can listen and tamper with any communications or computations. For example, the attack may come from a compromised operating system kernel or a malicious staff member, both situations in which the attacker has full control over the operating system. \algname guarantees the confidentiality, integrity, and provenance of any data in execution and in transit. 
The trusted computation base (TCB) of \algname is limited to the processor chip, codebase, and the WASM runtime running in an enclave, which explicitly excludes the operating system managed by the cloud provider. \algname does not guarantee against side-channel attacks, given that Intel SGX suffers from various side-channel vulnerabilities~\cite{10.1145/3052973.3053007, sgxsdk, tsgx}. Various techniques~\cite{10.1145/3052973.3053007,216033, tsgx,10.1145/2897845.2897885} proposed to mitigate the risk of side-channel attacks for enclaves.  

Since any entity outside the TEE can be malicious, we developed our consistency protocol to work in an asynchronous and Byzantine environment.
A network adversary can arbitrarily drop, delay, or replay packets.
Workers in our system can be put in network partitions for a finite but arbitrarily long time.
Our consistency protocol does not rely on timeouts for any action (e.g., view changes in consensus terminology) as a malicious OS can forever hold the system's progress in a livelock state.

We also assume the existence of a collision-resistant hash function (e.g., SHA256), a secure digital signature (e.g., Ed25519), and an authenticated encryption scheme (e.g., AES-GCM)

\subsection{System Overview}

\Cref{fig:overview} shows the components of \algname.
The FaaS Manager is responsible for launching FaaS workers on users' requests.
It has Key Distributor running inside an enclave which is responsible for securely distributing user's keys and inputs to the FaaS workers using attested TLS channels.
The FaaS worker contains a WASM Runtime, which runs JIT-compiled WebAssembly functions submitted by the user, with the securely sent inputs.
Multiple FaaS workers store their recent state updates in an in-enclave buffer called Memtable.
We build a \textbf{Secure Concurrency Layer (SCL)} that encrypts and signs the 
Since by assumption, messages can be lost in the network, we have a special long-running FaaS worker, the \Seq, which makes sure every worker sees linearizable updates in state.

For durability, we assume the existence of \textbf{replicated network-embedded \dcs}.
Particularly, to tolerate $f$ independent failures, we require $n = 2f + 1$ storage servers.
We always store to a majority quorum.
Persistent data is always encrypted using the user's application key.
The user also stores the function binary and the inputs and retrieves the output from these replicated \dcs.

% In this paper, we present PSL which provides a JIT-based WAVM runtime for enclaves on top of a decentralized and eventually secure consistent state update system.
% The rest of the paper is organized as follows:
% in \Cref{sec:overview} we discuss an overview of the system with threat model, system model and key management,
% in \Cref{sec:psl} we discuss the design of our execution environment,
% in \Cref{sec:kvs} we discuss our consistency protocol,
% in \Cref{sec:impl,sec:eval} we provide details of our implementation and evaluation methodology,
% in \Cref{sec:relworks} we discuss related works and conclude in \Cref{sec:fut_work}.

\subsection{Key Management}
Every \algname user generates two keys:
(1) \emph{Application Encryption Key},
(2) \emph{Application Signing Key}.

After attestation (see \Cref{fig:design:faas:sequence}), the Key Distributor establishes TLS channels between it and the FaaS workers and the user.
The user sends these two keys through the TLS channel which in turn is sent to the FaaS workers.
The \dcs only get the public key corresponding to the Signing key which they use to verify the signature on each block they store.

If the signature algorithm supports, the Key Distributor can generate child keys using a Hierarchical Deterministic Wallet approach \cite{gutoski2015hierarchical} and send one child key to each worker.
This allows fine-grained control over key access.
The \dcs are given the parent public key which they can use to verify signatures from any child key.

% For durability, we assume the existence of replicated \dcs. Concretely, we require $n = 2f + 1$ \dcs to survive $f$ failures.
% Note that, these $n$ servers must be in different failure domains.
% As we discuss below, we limit the attack surface of a malicious \dc to Denial-of-Service only.

% We introduce a FaaS execution service that enables secure \emph{and} stateful execution
% for both cloud and edge environments, while at the same time being easy
% to use.  The FaaS workers, called \textbf{Paranoid Stateful Lambdas}
% (PSLs), can collaborate to perform large parallel
% computations that span the globe and \emph{securely} exploit resources
% from many domains.  See Figure~\ref{fig:overview}.  We provide an
% easy-to-use data model and automatically manage cryptographic keys
% for compute and storage resources. 

% \section{Design}

% \subsection{Overview}

% \subsection{Execution}
% \textbf{WASM support for runtime and isolation}

% \subsection{JIT Compilation}

% \textbf{Script Attestation}

% \subsection{State}
% \textbf{Interface} \eric{Define KVS Interface here, especially with multi-language support}

% \subsection{Consistency}

% \subsection{Security Analysis}

\section{PSL Executor}
\label{sec:psl}

\subsection{Transitive Secure Worker Initialization} 
\label{sec:design:faas}
% \eric{a section better title}

\begin{figure}
    \footnotesize
    \centering
    \includegraphics[width=\linewidth]{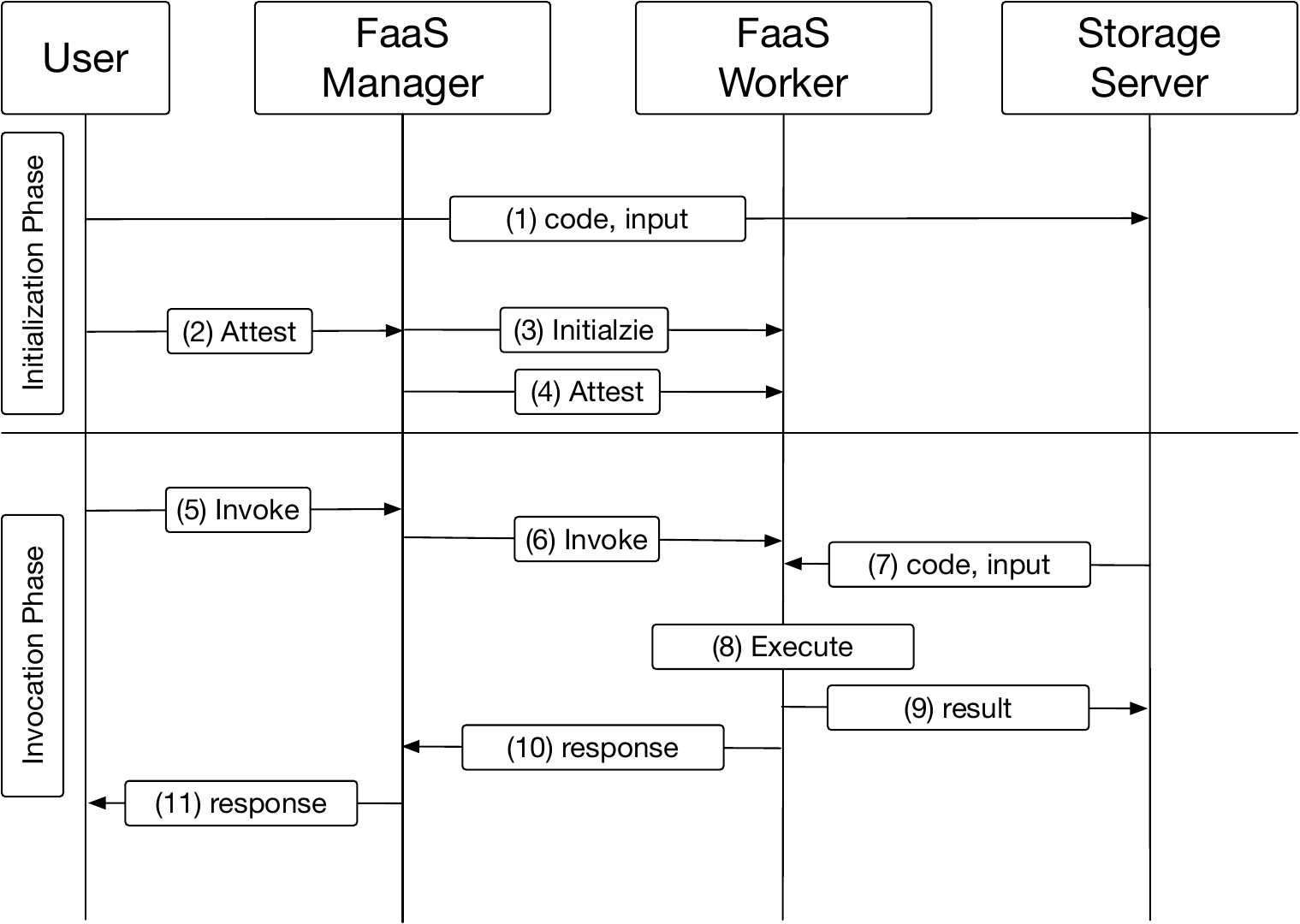}
    \vspace*{-0.2in}
    \caption{\textbf{FaaS Manager Launch Sequence Diagram:} At initialization time, (1) user securely uploads the function and input to \dc. (2-4) user attests FaaS manager while FaaS manager initializes and attests FaaS workers. When user invokes function, (5) they send a request to FaaS manager and (6) FaaS Manager distributes the request to other workers. The workers pull from \dc and execute the function. (9-11) On completion, the worker uploads the result to \dc and the FaaS Manager routes the response back to the user.  
    The detailed description of the Launch Sequence Diagram and the security protocol can be found in Section \ref{sec:design:faas} with the same step indexing. }
    \label{fig:design:faas:sequence}
\end{figure}

% \paragraph{Indirect FaaS worker Initialization}
Conventional secure enclave requires users to directly attest the remote machine and establish a secure connection between the user and the remote machine. This prevents man-in-the-middle attack. However, FaaS is a different compute paradigm that the workers should be pre-initialized before users issue any request. Then users can enjoy fast launching time and call the functions executed on remote machines without concern for infrastructure setup. As a result, we introduce a FaaS manager, an indirection proxy between workers and user, that facilitates the FaaS initialization process. 

Figure \ref{fig:design:faas:sequence} shows
%the workflow of using FaaS Manager to securely attest the application and 
the lifecycle of an application in \algname: 
\begin{enumerate}[wide, labelwidth=!, labelindent=0pt, label=(\arabic*)]
\item At initialization phase, user uploads the encrypted function code and function input to \dc with a key of hashed uploaded data. The code and function are encrypted with a symmetric key private to the client 
\item The client attests FaaS Manager with standard enclave attestation protocol to establish a secure TLS channel 
\item the client sends symmetric key to FaaS Manager through the secure channel 
\item FaaS Manager initializes and boots FaaS workers and their runtime environment. This step corresponds to typical FaaS frameworks which initiate a pool of active workers that stand by and await for requests. 
\item FaaS Manager attests FaaS workers and establish a secure TLS channel to each  worker
\item At run time, client invokes the function with the keys from \dc of the function and input 
\item FaaS manager finds idle workers, and send with the keys from \dc and the symmetric encryption key 
\item FaaS worker retrieves the input and function from \dc 
\item FaaS workers execute the function based on the input and exchange the intermediate results with other workers. 
\item At the end of the execution, FaaS worker stores the result in \dc and returns FaaS manager with the corresponding key. 
\item FaaS Manager return client with the key. Client retrieves from \dc and gets the final result.

\end{enumerate}

\paragraph{Sandbox Reloading} 
Loading and attesting each client function per invocation for an enclave would have an unreasonable initialization time. As a potential optimization, we argue for reusing the execution environment across worker functions. In this case, we save round trip time due to attestation and bootloading time for the enclave. How do we ensure that reusing execution environments across workers is safe? We propose using a secure sandbox to disallow client functions from hijacking the worker enclave. The FaaS manager is responsible for launching and pre-attesting worker enclaves to ensure that they run the correct sandbox. Once the worker enclave finishes execution, it maintains a memory array that was allocated and used throughout the worker's execution. The sandbox will force the worker to only write to memory within this memory array, which is freed and zero'd out after the worker finishes executing or aborts due to errors.

% The FaaS manager also runs a key distribution protocol to provide unique private keys (derived from client secrets) for each worker.  

\subsection{WASM JIT Runtime}
% \eric{section copied from poster, @Alex, please help with this section}

In section \ref{subsec:case_for_jit}, we argued for using a JIT runtime engine as our FaaS worker execution model. Implementing a secure JIT runtime engine is paramount. SGX1 did not allow for dynamic memory mapping, so in order to even allow a JIT runtime engine in SGX1, you needed to allocate a sufficiently large, static, and executable reserved section, which would be where the generated code resides. First, with the heteregenous workloads and memory consumption of FaaS, it isn't sufficient to have a static memory reservation for all workers. Second, as the developers cannot call mprotect on the reserved section, they risk having a code injection attack that might hijack the sandbox. 

Thus, our solution for our runtime engine should satisfy the following properties 1) the amount of memory that the runtime engine allocates for the worker should be dynamic 2) and the runtime engine should have the ability to unmap or change the permissions of pages dynamically.
% \paragraph{Dynamically Remapping Pages in SGX2}
Enclave dynamic memory mapping (EDMM) support was introduced in SGX2, which enables mapping and modifying the permission of pages dynamically after the enclave has been measured. We use EDMM to dynamically allocate memory for each client as well as unmap and clear the client's memory when execution is completed. 

We port WAVM \cite{wavm}, a WASM JIT runtime engine that relies on the LLVM framework for code generation, inside of SGX. We modify WAVM to use EDMM to mmap buffers in-enclave for the client. The WAVM runtime has a loader which fetches code keyed on a hash outside the enclave. Once the code is fetched, the runtime will verify that the code is correct by hashing the module and matching it with the original hash. Once the code is verified, it is mapped and loaded, then goes to the LLVM pipeline. This involves emitting LLVM intermediate representation (IR), optimizing the IR, then generating the machine code. The memtable is protected and managed by the runtime, so that potentially malicious clients are sandboxed from accessing memtable state that it doesn't own. 

The JIT runtime also links in our own WASM module that interfaces with the KVS with simple ReadKey/WriteKey interfaces. Clients simply need to include a header file with no other dependencies. On top of the KVS interface, we create shared memory through familiar array abstractions that we call PSLArray. Multiple workers can easily share data by writing to a PSLArray, where updates are transparently multicasted to other workers. Later, we demonstrate how one can use a PSLArray to have a shared distributed weight array that multiple workers read and write to for deep learning training.

% Loading and attesting each client function into an enclave would have an unreasonable initialization time. We instead add a level of indirection, where clients contact a trusted FaaS manager that is running inside an enclave that the client can attest. The FaaS Manager is responsible for launching and attesting PSL worker enclaves running a WASM runtime. The FaaS manager will only have to attest that the correct WASM runtime is loaded in the enclave (the certifier component in Figure~\ref{fig:overview}). The FaaS manager also runs a key distribution protocol to provide unique private keys (derived from client secrets) for each worker.  

\section {Secure Concurrency Layer (SCL)}
\label{sec:kvs}

In this section, we formally define our consistency guarantees and system constraints with our system design.

To the application running in a FaaS worker, our state management mechanism interfaces as a shared memory key-value store.
We define three key guarantees of our system:
(1) Monotonicity,
(2) Eventual Progress,
(3) Validity.
We define these formally below:

\begin{definition}{\textbf{Monotonicity.}}
\label{definition:monotonicity}
If a worker reads value $V_1$ for some key $k$, all subsequent reads for the same key $k$ from the same worker returns values $V_2$ such that
$$V_1 \leq_{e} V_2$$
for some partial order $\leq_{e}$ on the space of values.
\end{definition}

% With Definition 1, all subsequent reads are done by the same worker.
% Unlike a distributed database, where external clients need to see linearizability from the system as a whole, a FaaS system has applications tied to the worker they are executing on.
% Hence, Definition 1 of linearizability fits FaaS.

\begin{definition}{\textbf{Eventual Progress.}}
\label{definition:progress}
If a worker writes value $V_1$ for some key $k$, every worker in the system will eventually read values $V_2$ for the key $k$, such that,
$$V_1 \leq_{e} V_2$$
\end{definition}

Note that, we operate in an asynchronous environment where packets can be delayed or dropped arbitrarily.
Eventual progress guarantees that updates made by a worker are visible to all other workers eventually.
However, it does not provide time bound for the progress to be visible.

\begin{definition}{\textbf{Validity.}}
If a worker reads value $V$ for some key $k$, there exists a worker that wrote the same value $V$ for the key $k$.
\end{definition}

This property guarantees that no network adversary can maliciously inject state updates into the system.

% Discuss constraints and insights
% Constraints: Limited memory; remote storage; everything is encrypted.
% Insights: State is more important than logs; locality; passive replication
We observe the following constraints in a TEE-based FaaS system:
Firstly, TEE platforms often come with limited available memory. For example, Intel SGX has an Enclave Page Cache (EPC) size of only 128MB. The encrypted paging becomes the performance bottleneck for large applications. Hence, we must be efficient in the in-memory cache usage of our application. Note that this is different from the current design of systems like CCF~\cite{russinovich2019ccf}, which assumes that the application state is completely in memory.
Secondly, we do not assume that the TEE-enabled machines have large disk capacities. We make this distinction due to our heterogeneous infrastructure assumption. Thus, all access to persistent data must be through network-attached \dcs.
Thirdly, for confidentiality, all data sent outside an enclave should remain encrypted. This incurs additional costs of encryption on the critical path of sending messages. Also, queries on data cannot be performed outside the TEE.

Existing systems gravitate towards running a consensus protocol among the FaaS workers.
This makes these systems perform active replication: every worker first agrees on a common total order of commands or transactions and then executes them to update their local state.
We argue that, if the application does not need a strict total ordering of its commands, active replication under-utilizes the guarantees a TEE provides.
A trusted environment guards against malicious code execution.
If the application logic (provided by users of our system) is correct, multiple FaaS workers could be running multiple transactions at the same time and merge the results eventually.
This model of \textit{passive replication} is more suited for a TEE-based environment.

% Discuss eventual update rule: why not vector clocks per key? Cache update rules
\subsection{Merge Operation}
With the above constraints in mind, we now discuss our system design for eventual consistency.
Central to this discussion is the partial order $\leq_{e}$ used for merging two values for the same key.
We assume each value $V$ is structured as $V = (v, ts)$ where $v$ is the actual data and $ts$ is a timestamp attached to it. We define the partial order as follows:
$$V_1 \leq_{e} V_2 \iff (ts_1 < ts_2) \vee (ts_1 = ts_2 \wedge H(v_1) \leq H(v_2))$$
where $H(.)$ is a collision-resistant hash function.

Many eventually consistent systems \cite{dynamodb} use vector clocks for their timestamps to capture causality, but the size of vector clocks is linear to the number of workers.
Due to memory limitations in our system, we did not use a vector clock, rather, we used a Lamport clock. This works because the only communication between our FaaS workers is through this eventually consistent key-value store system.
When timestamps are equal, we use the hash of data to break the tie instead of node identities. This mitigates the security problem that a network adversary may influence the result of the merge by selectively dropping packets from workers.

% Commit guarantees durability.
\subsection{Durable Writes}
Every FaaS worker has a local cache of the writes in the system, called the \textbf{Memtable}.
To the application worker, we expose a transactional interface that publishes a batch of writes at a time.
% While these transactions are not ACID \cite{something_acid?},
We only guarantee durability on commit.
% \footnote{We are consistent with BASE transactions; common for NoSQL databases. \cite{aws_article}}
For every key-value update, the Memtable increases the Lamport timestamp by one.

\begin{figure}
    \centering
    \includegraphics[width=\linewidth]{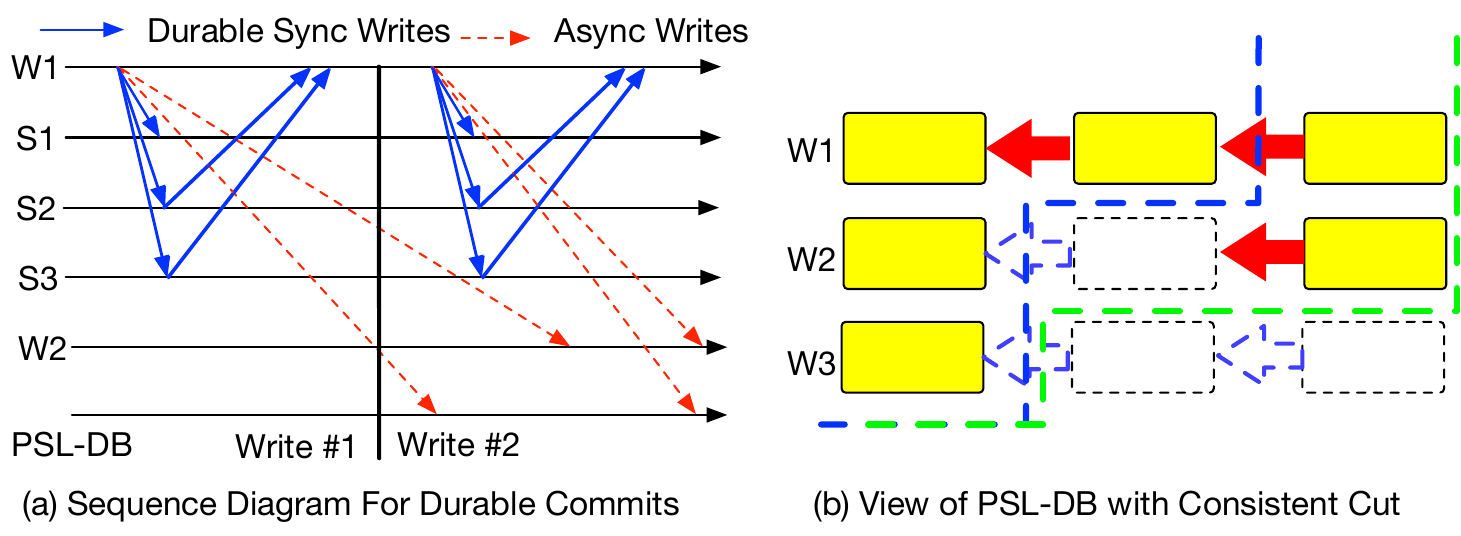}
    \caption{(a) \textbf{Durable Commits.} Worker $W_1$ multicasts a batch of write to \dcs ($S_1, S_2, S_3$), other workers ($W_2$) and \Seq. It only waits for f + 1 responses from \dcs and progresses to multicast the next batch. (b) \textbf{View of \Seq.} It has all the writes from $W_1$. It lags behind $W_3$ and it has missed an intermediate write from $W_2$. The blue line shows a causally consistent cut. The green line is not a causally consistent cut until the \Seq back-fills the missing block from $W_2$.}
    \label{fig:consistency_time}
\end{figure}

\Cref{fig:consistency_time} describes the operations that take place on Commit.
We assume the existence of $n = 2f + 1$ \dcs, where at most $f$ can fail.
The replication on this set of \dcs is controlled by the FaaS worker itself.

In the FaaS worker, whenever the application performs a write, the writes are all stored in a transaction buffer.
Once the application calls Commit on the transaction, these writes are first stored in the Memtable and then multicast to every other worker in the system. Simultaneously, these writes are sent to all the \dcs and the worker waits for $f + 1$ of them to return an acknowledgment.
The commit completes once the acknowledgment arrives.
We attach the hash of the previous multicast block to the current block of writes and give it an increasing sequence number. Then the block is encrypted using an authenticated encryption scheme before sending it out.
The \dcs only see the encrypted blocks, which are also periodically signed by the FaaS worker.

Once other workers receive a multicast block, it applies all the writes to its own Memtable. If the keys already exist, it uses the partial order $\leq_{e}$ to merge the values. 
Other workers can miss some writes due to packet drops, but that does not break the monotonicity guarantee.

% Sequencer does causal consistency.
\subsection{\Seq Consistency Manager}
% We have a special FaaS worker, the \Seq, for every application that we keep alive for the entirety of the application's lifetime.
\Seq periodically generates global snapshots of all writes made by all the FaaS workers. It facilitates workers to recover missed messages and acts as a database for keys not present in the workers' Memtables. It needs to be alive for the entirety of the application's lifetime. 
% Since we allow other FaaS workers to miss messages due to packet drops, we use the \Seq to periodically generate global snapshots of all writes made by all the FaaS workers.
% Additionally, it also 

The \Seq receives the multicast messages from workers and keeps track of their sequence numbers to detect missing blocks.
If a missing block is detected, the \Seq uses hash pointers attached to the later block it received to back-fill the missing block from \dcs.
It also periodically requests all \dcs to return the most recent block sent by each worker and waits for the response from $f + 1$ \dcs.
The block returned with the maximum sequence number is guaranteed to be the most recent block from the worker.
If this sequence number is higher than the one the \Seq has seen, it uses this block and its attached hash pointer to back-fill any block that it might have missed.

This ensures that the \Seq's view is always a causally \textbf{consistent cut}\cite{chandy-lamport} of all writes by each worker (i.e., the subset of all writes seen by \Seq at any point is such that for all write blocks in the set, its attached hash pointer points to another write in the set, or is null).
Due to message drops, it may be lagging behind the most recent writes.
% but that does not violate linearizability for individual key-value pairs.

The \Seq has its own Memtable which is used to merge all key-value pairs in its received blocks.
It periodically flushes the Memtable out into a Checkpoint block which is also stored in a quorum of \dcs.
Then it compresses the Checkpoint into a \textbf{Sync Report} which contains the list of keys and their corresponding timestamps and hash of the value stored in the checkpoint and the most recent sequence number seen from each worker.
This Sync Report is then multicast to all the workers.
The workers use the Sync Report to query the keys that are more updated in \Seq than their own Memtable.
The Sync Report also contains an increasing sequence number and a hash pointer to the previous Sync Report and is replicated in $f + 1$ \dcs. So, a worker can also recover and back-fill missed Sync Reports from \dcs using the hash pointers.  
 \Seq can also answer queries for keys, either returning directly from its Memtable, or sending the hash of the Checkpoint block that has the most recent value for the key.
 % If a worker misses a few intermediate Sync Reports, it back-fills them from \dcs using the hash pointers.

% As stated before, \Seq also answers queries for keys, either directly from its Memtable, or sends the hash of the Checkpoint block that has the most recent value for the key.

\subsection{Release-consistent Locking}

% The above protocol works correctly for applications that don't need a global total order over its operations.
% For applications that do need it, we provide a way for workers to perform reads and writes mutually exclusively of each other using locks.
Some applications require additional mutually exclusive reads and writes using locks. \Seq implements \emph{release-consisent locking}~\cite{gharachorloo1990} with a Lock Manager that grants locks on worker's requests.
% The locks are granted on request to workers by a Lockserver attached to the \Seq.
Once a worker releases a lock, it attaches the hash of its most recent write block with a lock release message to the Lock Manager.
The \Seq, on receiving the message, makes sure it has seen that block and then issues a Sync Report.
The Sync Report is also attached with the lock grant message to the next worker which requests to acquire the lock, thereby making sure that that worker is updated with all the writes made by the previous writers.
This makes the locking system release consistent and allows the application to have total order, albeit at the cost of performance.

\subsection{Discussion}
\noindent \textbf{Memtable size.}
% While this is not a requirement for correctness, for performance reasons, any application running in our FaaS worker should have the Memtable big enough to fit the working set of keys to avoid thrashing.
To avoid thrashing and improve application performance, the size of a Memtable should fit the working set of keys. 
We provide a low cache mode if writes from other workers can evict a key in the cache before being included in a Sync Report: 
% When the cache size is so low that multicast messages from other workers can evict a key before it is included in a Sync Report, we offer a low cache mode, which does the following:
(1) Pins keys not included in a Sync Report to the cache. Any attempt to evict these keys from the Memtable blocks until a Sync Report with this key included arrives.
(2) Does not let multicast blocks from other workers introduce new keys in the Memtable.
This preserves linearizability even when a recently written key is forced to be evicted due to low Memtabl size.

\noindent \textbf{Garbage Collection.}
Note that, the blocks made durable in \dcs by the workers serve as a Write-Ahead Log (WAL) for the \Seq.
Once the \Seq puts the writes of these blocks into a Checkpoint, all future reads happen from this Checkpoint block only.
Hence, once a block is added to a Checkpoint, it can be garbage-collected from the \dcs.

\noindent \textbf{Behavior under failures.}
We assume that the FaaS workers have a crash-stop failure model. We don't allow a restarted FaaS worker to reuse its old identity.
Note that, \Seq is \emph{not} a single point of failure in the system.
Although we do all our evaluations with a single \Seq, multiple \Seqs could be active at any given point in time, without a need for coordination among them.
In that case, the workers need to send their key-value pair queries to the \Seq from which they received the most recent Sync Report.
A \Seq can crash mid-operation and be restarted directly.
The back-filling strategy triggered by multicast from the workers will make sure that before the next Sync Report, the \Seq has seen all the blocks from all the workers.
While the \Seq is in a crashed state, the workers still keep making progress using the multicast mechanism.
However, since we consider an asynchronous network environment, all queries to the \Seq for key-value pairs and all lock requests will be blocked until the \Seq comes back up and catches up with all the workers.

\subsection{Correctness}

We state the following lemmas:

\begin{lemma}
\label{lemma:linearizability}
Monotonicity with eventual progress imply linearizability.
\end{lemma}

\begin{lemma}
\label{lemma:correctness}
Under an asynchronous network environment, the protocol described above provides linearizability, eventual progress, and validity.
\end{lemma}

% The proofs are left in the Supplementary material for brevity.
We omit the proofs here for brevity.
\section{Implementation}
\label{sec:impl}
We implement PSL with $\sim$ 21k lines of C/C++.
We use a fork of the OpenEnclave framework for our Intel SGX-specific code that supports EDMM. In order to port WAVM, we ported the LLVM libraries to link with OpenEnclave's framework.
We create point-to-point persistent TCP channels using ZeroMQ.
Communication among FaaS workers, the \seq and \dcs happens using an RPC format.
% The exact specs of the RPCs are defined in Algorithms \ref{alg:dc_op}, \ref{alg:worker_op}, and \ref{alg:seq_op} in the Supplementary.
We use Ring Buffers \cite{openenclaveswitchless, weisse2017regaining} to asynchronously send messages to enclave threads.
This circumvents the need for Ecalls for every message.
The implementation of our FaaS worker can scale up to use each core available in a machine effectively.
We use one server thread to receive messages and two for processing app launch, key exchange, multicast, and Sync Report messages.
The rest of the cores can be used to run the application.
The Memtables for FaaS workers are implemented using LRU caches.

The \dcs are implemented using RocksDB for persistence.
We disable write-ahead logging and use an 8GiB write buffer size for performance reasons.
We also provide two alternative designs, one directly using the underlying file system (e.g., ext4), and another providing a shim layer over a cloud storage service (e.g., Azure Blob Storage).
We noticed that for blocks smaller than 4MiB, RocksDB provides stable and high write throughput, as the blocks are mostly cached in memory.
For larger blocks, the file system design is better since it frequently fills the write buffer in and causes unnecessary compactions in RocksDB.
Since most of our experiments generate blocks smaller than 4MiB, we use the RocksDB system for our evaluation.
The cloud storage version of the \dc provides high-latency operation. 

For authenticated encryption, we use AES-GCM with 256 bit keys.
AES-GCM generates a small tag containing the Galois MAC of the data encrypted.
This tag is sent along with the encrypted data itself.
Anyone holding the symmetric decryption key can verify the integrity of the encrypted message using this tag.
We use RSA with 2048 bit key size for key exchange with the FaaS manager.
For all digital signatures, we use Ed25519.
We previously used seckp256k1.
Whereas secp256k1 provided more fine-grained control over key management by using a child key derivation scheme like Hierarchical Deterministic Wallet \cite{bip32}, Ed25519 has an order of magnitude better performance in signing and verification.
We use SHA256 as our collision-resistant hash function.
All crypto operations are programmed in OpenSSL3.0.

We mention a few notable optimizations below.

\subsection{Runtime Optimizations}
% \shubham{@Alex: Probably put some more?}
\paragraph{Reducing Startup Costs}
In order to reduce start-up latency, we first identify the the phases that account for the bulk of the cold start up latency 1) allocating memory for the worker and 2) generating machine code. We predict and pre-allocate memory to hide the latency required during cold start up time for each worker. To prevent machine code generation, we cache modules generated by JIT. We further allow users to AOT compile binaries to save on cold start latency costs. We demonstrate how much speed-up we achieve through these optimizations in our evaluation section.

\paragraph{Concurrency}
We enable a promise-based concurrency model in our system.
The main application thread can run any function asynchronously and get a promise, which it can use to wait for the result later on.
If there is only one application thread available, promises behave as if they are run synchronously.
However, if more than one application thread is available, the promises are handed to idle threads for parallel execution.

% We use the Web Assembly Virtual Machine (WAVM), which uses LLVM to dynamically compile to WASM bytecode. We link the LLVM libraries with a modified version of OpenEnclave, which supports dynamic memory remapping. We are actively porting WAVM to run in the OpenEnclave framework. We also plan to work on a port to RISC-V
% % ,  using the Keystone TEE framework \cite{keystone}, 
% for better exploring computation on the edge.

\subsection{SCL optimizations}
\paragraph{Multicast Batching}
Since the Memtable is shared among application threads and the multicast thread,
updating the Memtable with updates from multicast messages requires acquiring a lock for mutual exclusion.
To reduce lock contention, it is preferable to have the update time short.
After every iteration of the multicast updates, we check if there are more than one multicast updates in the Ring Buffer.
If so, we collect all of them together and batch them into one big multicast block.
In the next iteration of updates, we use this big block for updates.
This eliminates redundant keys in the original batch. Each key only appears once in the big block.

\paragraph{\Seq LSM tree}
Since \Seq is also used as a database for key-value pairs, requesting a key from \Seq should be made fast.
However, \Seq also runs in a limited memory enclave and has network-attached \dcs only.
Under these design considerations, we adapt the Log-structured Merge (LSM) tree approach taken by databases like RocksDB and LevelDB.
We implement the Memtable using C++ maps.
When the Memtable is filled with a threshold number of keys, we flush the Memtable and cache the resulting checkpoint block in memory.
A fixed-length list of cached checkpoint blocks makes the level-1 in LSM tree terminology.
The level-2 consists of a sorted set of keys.
The keys are divided into ranges, where the \Seq has to keep the first key of each range in its memory.
The rest of the range is durably stored in \dcs and is also held in another cache.
When level-1 fills up, a constant fraction of the blocks are compacted into level-2 and then erased from memory.
This is where the design differs from traditional LSM trees.
Whereas generally, LSM trees fetch data from disk to perform compaction, we maximally utilize the in-memory cache by performing the compaction eagerly in memory.
The sizes of all these levels can be configured.
For fetching keys, we first search the Memtable, then the level-1 blocks in descending order, and then finally level-2.
The keys returned from level-1 and level-2 do not contain the actual values. Rather, they contain the hash of the Checkpoint block that has the value.
This is done to prevent double buffering. The FaaS worker requesting the key should, for locality, fetch the Checkpoint block and apply all its writes to its Memtable.

\section{Evaluation}
\label{sec:eval}

\begin{table}
    \vspace*{0.1in}
    \centering
    \footnotesize
    \begin{tabular}{l c c  }
      \toprule
      & Median  & $99.99^{th}$ percentile  \\
      Operation  &  latency ($\mu$s) & latency ($\mu$s) \\ 
     \midrule
     ICMP Ping & 641.5 & 1337.7 \\
     % \hline
     Writing to quorum & 1000 & 180000 \\ 
     % \hline
     AES-GCM on 2KiB block & 2.0 & 39.1  \\
     % \hline
     Ed25519 Sign & 44.0 & 61.0 \\
     % \hline
     Ed25519 Verify & 150.0 & 745.3 \\
     % \hline
     RocksDB write & 7.0 & 357.2 \\ 
     \bottomrule
    \end{tabular}
    \caption{\textbf{\algname Cluster System Characteristics}}
    \label{tab:environment}
    \vspace*{-0.25in}
\end{table}

\begin{figure*}
    \centering 
    \includegraphics[width=0.9\linewidth]{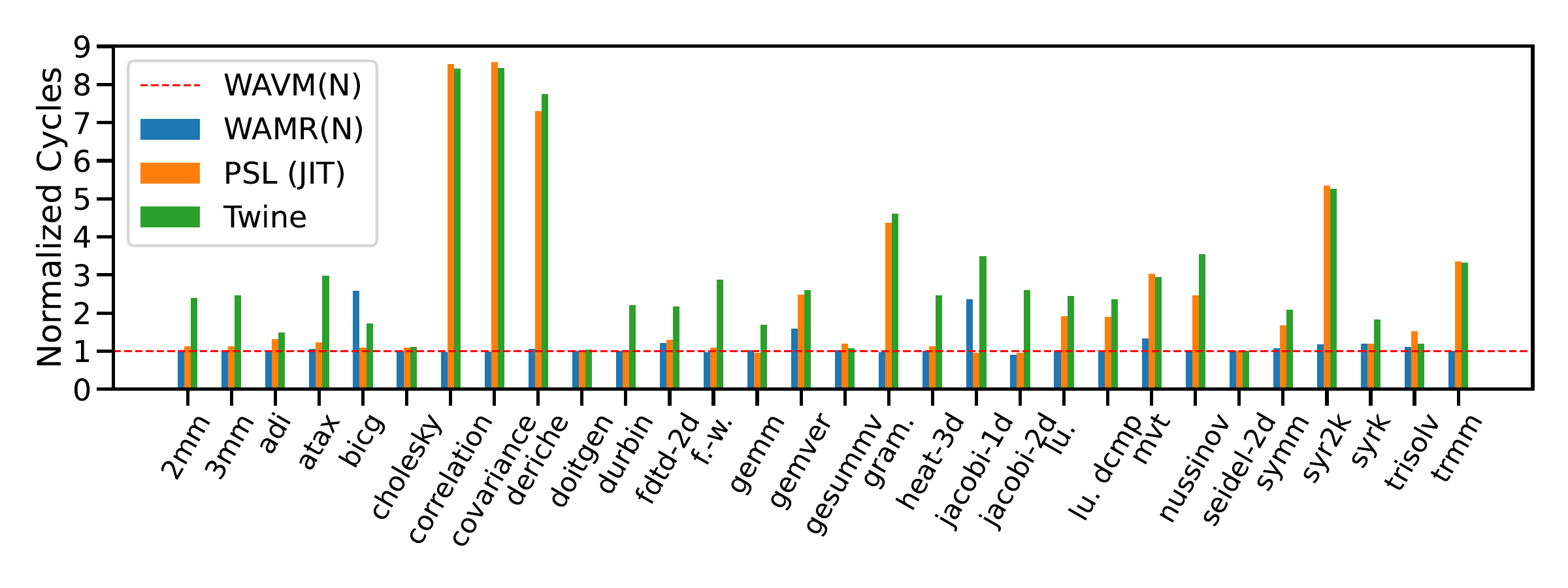}
    \vspace{-20pt}
    \caption{\textbf{Execution Cycle Comparison on PolyBench between Twine\cite{menetrey2021twine} and \algname. }  All cycles are normalized by running PolyBench on WAVM (Native) without SGX. WAMR (the runtime of Twine) and WAVM (the runtime of \algname) demonstrate similar Native performance. However, \algname-JIT demonstrates up to 3.7 times latency improvement compared to Twine.}
    \label{fig:polybench}
\end{figure*}

\subsection{Evaluation Setup}
We use Azure DCds\_v3 series machines with Intel(R) Xeon(R) Platinum 8370C CPU @ 2.80GHz and Linux kernel 6.5 for our Intel SGX2 environments.
For our experiments on SCL, we use a cluster of 10 machines with 8 vCPUs and 64GiB memory. For our experiments on the runtime, we use 2 machines with 16 vCPUs and 128GiB memory.
For \dcs, we use three Azure D8d\_v5 machines with 8 vCPUs and 32GiB memory.
We use a partition of 128GiB ephemeral storage.
We set up a Kubernetes cluster using K3S with the SGX2 and storage machines where we set the resource requests in such a way that each entity (FaaS manager, \Seq or worker) gets a machine by itself.

\subsection{Execution Microbenchmarks}
In \Cref{fig:polybench}, we compare with another SGX runtime Twine \cite{menetrey2021twine}, which uses AOT compilation and is based off of the Web Assembly MicroRuntime (WAMR). We run the Polybench \cite{ReisingerPolyBenchC} suite which is a collection of compute bound benchmarks compiled using Emscripten \cite{emcc} with -O3. For Twine, the benchmarks were compiled using WAMR's own AOT compiler (wamrc) with -O3. We also compare PSL against running the benchmarks on WAVM outside of SGX. PSL outperforms Twine in a majority of the benchmarks, even achieving up to 3.7x speedup over Twine. In order to ensure that the performance is not simply due to the differences in runtime, we normalize the performance relative to running WAVM outside of an enclave and compare against WAMR, the runtime Twine uses. We see that the majority of benchmarks WAVM and WAMR perform around the same, yet PSL's performance over Twine remains strong.

Compared to native performance, we expect slowdown as previous work has shown that running in WASM generates more memory operations, produce more instructions that branch, and generates more instructions \cite{234914}. A cache miss in SGX is costly, relative to outside an enclave, as the memory has to be decrypted when it goes into the processor. For example, the correlation benchmark of Polybench does operations on > 20 MiB, which is the size of the L1 + L2 cache combined. This accounts for why 2 of the benchmarks, correlation and covariance have a significant slowdown due to L1/L2 cache missing. Note that the same slowdown also appears in Twine.

\begin{figure}
    \centering
    \includegraphics[width=0.9\linewidth]{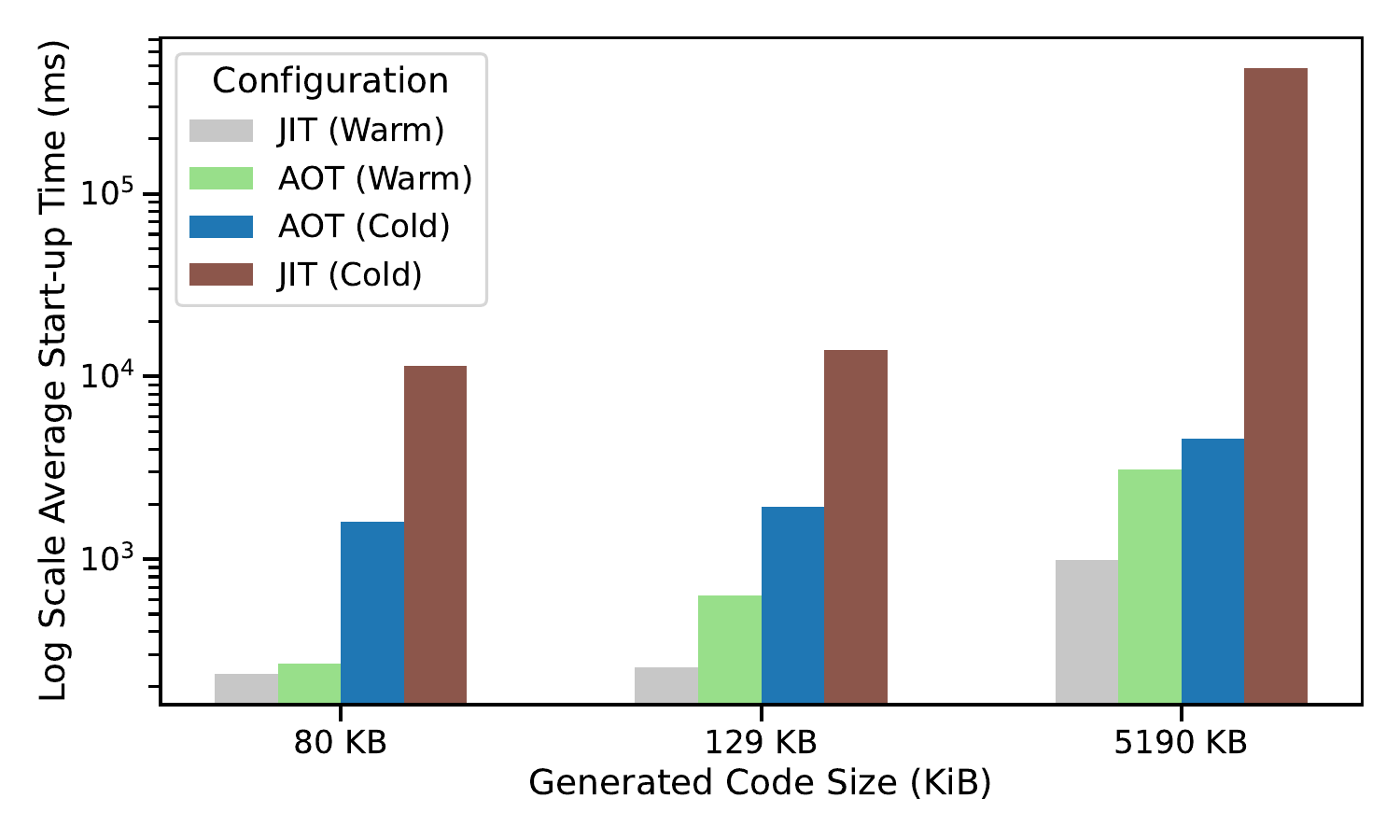}
    \vspace*{-0.15in}
    \caption{\textbf{Startup Time Comparison of PSL Under different WASM Compilation Options and Cache Status}  The startup time includes the time of fetching the code and input from \dc and loading the code to enclave.  }
    \label{fig:eval:startup}
    \vspace*{-0.05in}
\end{figure}

\subsection{Startup Latency}
In Figure \ref{fig:eval:startup}, we show the performance of an active worker retrieving a invocation request up until it returns the first instruction. Cold start up latency for JIT scales with the binary code size as generating machine code dominates cold start latency. As expected, for cold AOT, the start up doesn't scale as much as JIT with binary size. For both JIT and AOT, caching the generated code significantly improves performance. For PSL workers, this demonstrates the importance of pre-fetching and warming up the cache with libraries we expect the user to link with.

% \eval{execution latency (single-worker) on X applications, compared to graphene, wamr}

% \textbf{Insight: WASM by \algname is fast by XXX times compared to in-enclave containers, and native JIT is fast compared to reinterpreted } 

\subsection{SCL Benchmarks}
\paragraph{Workloads}
Unless stated otherwise, we use the YCSB\cite{cooper2010benchmarking} benchmark to test SCL.
We use a workload of 300,000 unique keys with 100 byte values.
Since we are operating in a FaaS environment, we run YCSB from trace files, rather than using a client machine.
Workload trace files with varying read-write ratios are generated per worker and stored in the \dcs.
The workers are given the hash of the trace files as input.
The workers retrieve its file from \dcs and start processing requests in batches.
We use a batch size of 20 in our experiments.
We let each experiment run for at least 5 minutes and sample throughput at every 5s interval once it is stabilized.

\begin{figure}
    \centering 
    \includegraphics[width=\linewidth]{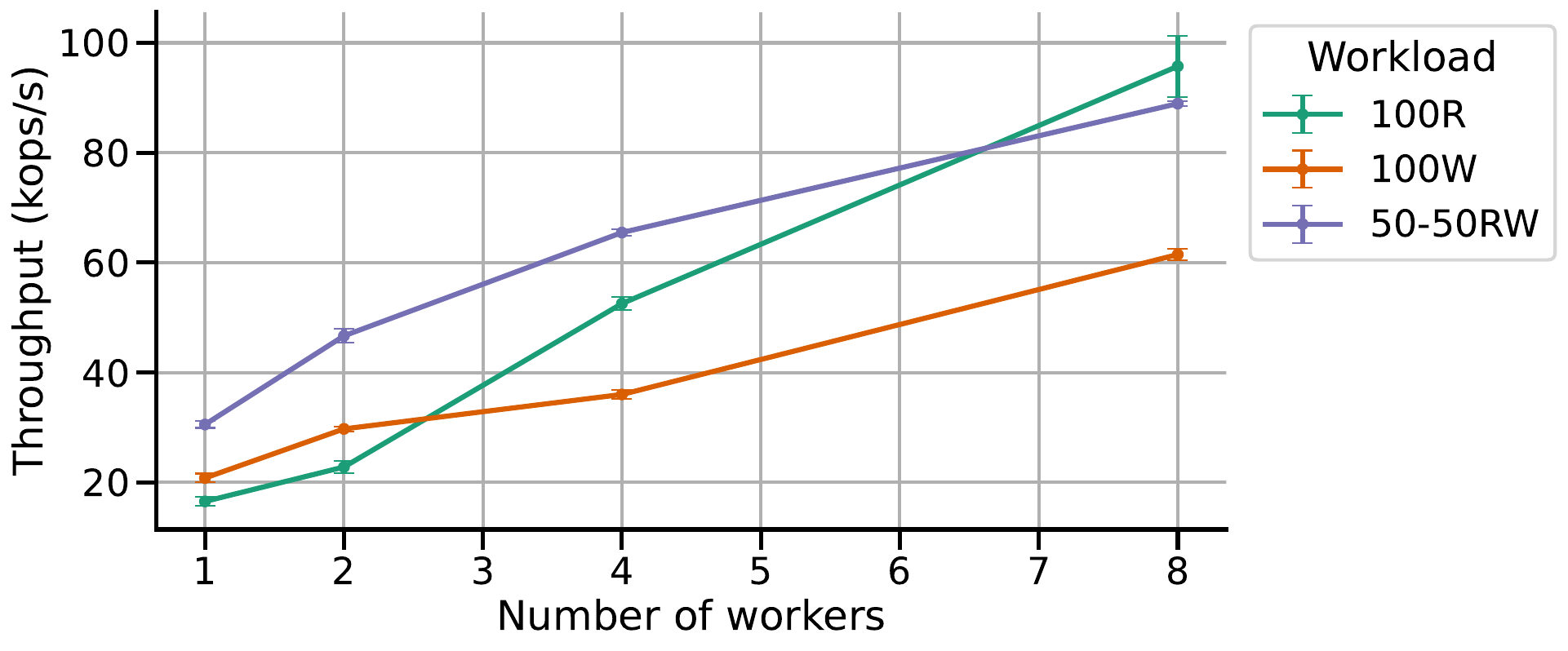}
    \vspace*{-0.15in}
    \caption{\textbf{Throughput Scalability With The Number of Workers on Different Read-Write Ratios} Each worker runs 6 application threads and signs for every 200$^{th}$ block.}
    \label{fig:scaling}
\end{figure}

\paragraph{Scaling}
We run the YCSB workload with 0, 50, and 100\% write ratio. \Cref{fig:scaling} shows the scaling of the system for 1, 2, 4, and 8 FaaS workers running 6 application threads each.
We achieve up to 95k operations per second for the 100\% read case.
The 100\% write case does not scale as well due to multicast traffic.

\begin{figure}
    \centering 
    \includegraphics[width=0.75\linewidth]{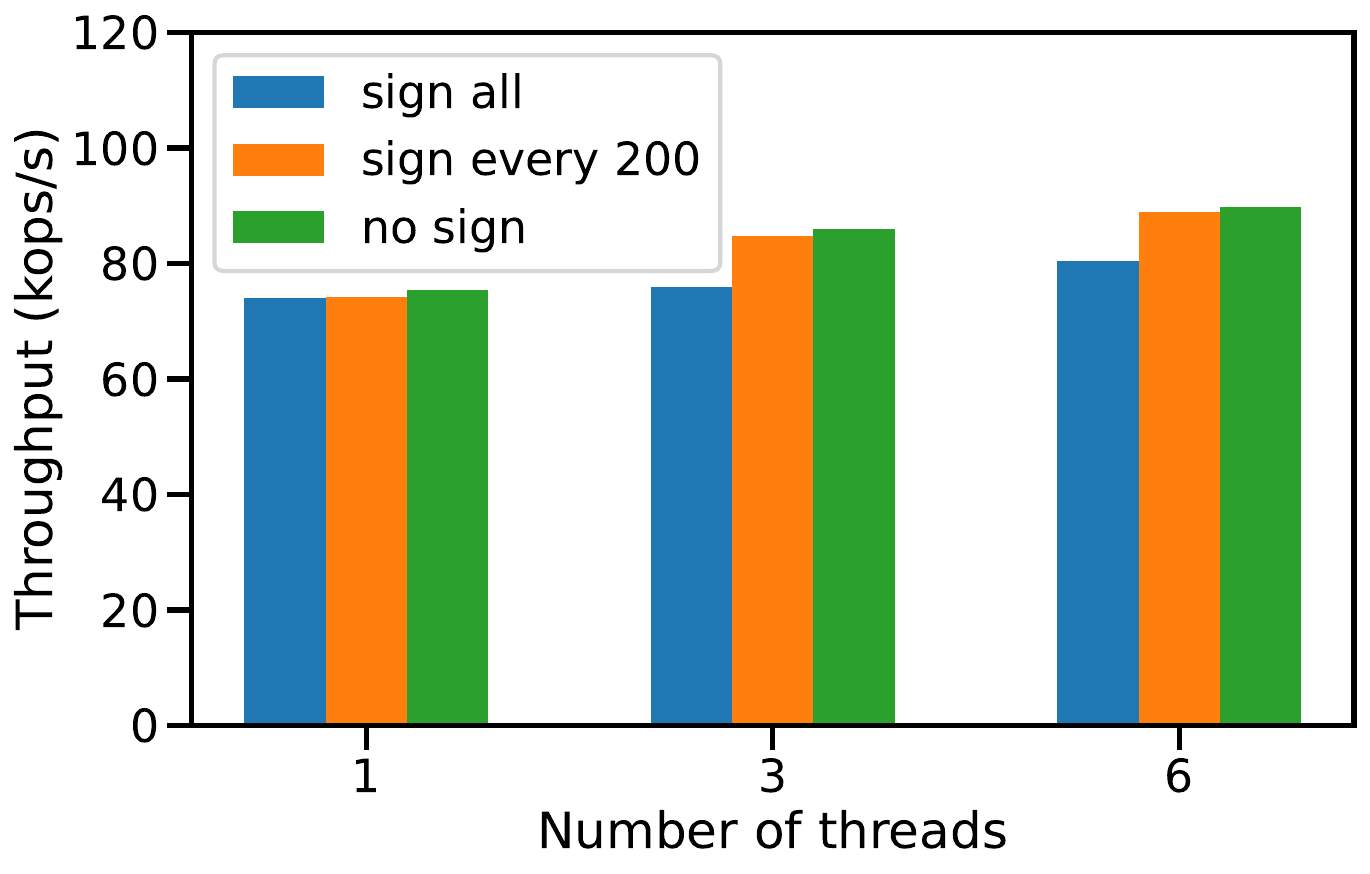}
    \vspace{-10pt}
    \caption{\textbf{Overhead of Signatures vs Number of application threads in an 8 worker setup.} "sign all" signs every block; "sign every 200" defers signature for every 200 blocks; "no sign" turns off both signature and verification, and only encrypts the blocks. The throughput overhead from signing all blocks to not signing is $\sim$10\%.}
    \label{fig:crypto}
\end{figure}

\paragraph{Cryptographic overhead}
From \Cref{tab:environment}, we know that the individual cryptographic operations take 10s $\mu$s.
We run an eight worker setup with number of 1, 3, and 6 application threads and vary the amount of signatures done.
\Cref{fig:crypto} shows the aggregate throughput achieved by the system.
The overhead of signatures is $\sim$ 10\%.
The system saturates the network as well as the CPU cores in the machine.

\paragraph{Release-consistent locking}
% \begin{figure}
%     \centering 
%     \includegraphics[width=\linewidth]{figures/lock.pdf}
%     \caption{\textbf{Caption}}
%     \label{fig:lock}
% \end{figure}
Release-consistent locking mechanism gives a considerable overhead.
The throughput of a single-threaded 1 worker system with no locking is around 17k ops/s. However, that with locking drops to 8k ops/s.
This throughput remains almost fixed as we scale to 2, 4, and 8 workers as mutual exclusion only allows one worker to progress at a time.
We therefore conclude that release-consistent locking should be used very rarely.
We emulate single-threaded Paxos in our implementation and the throughput is 9.2k ops/s which is close to our throughput with release-consistent locking.

\paragraph{Cache performance}
To understand the performance of the system with bigger application sizes, we increase the number of keys in the YCSB trace such the total YCSB application size becomes 256MiB.
For this application we run with Memtable sizes of 64, 128 and 256MiB.
We get throughput of 53.6k, 54k, and 63.2k ops/s respectively for a 50\% write workload.
Since SGX comes with 128MiB EPC size, using caches higher than that size does not scale well due to expensive paging operations.

\begin{figure}
    \centering 
    \includegraphics[width=\linewidth]{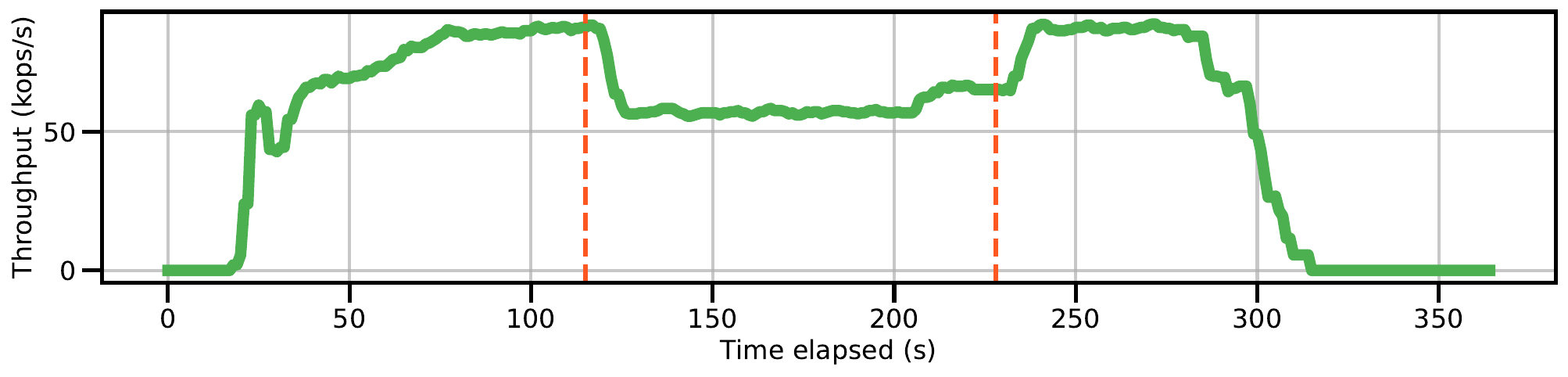}
    \vspace*{-0.2in}
    \caption{\textbf{Throughput Timeline under transient \dc failures.} The red lines denote the exact time one healthy \dc crashes and comes back up respectively. During the downtime the commits happen through 1 healthy and 1 lossy server, prompting multiple retries.}
    \label{fig:dc_crash_test}
\end{figure}

\paragraph{Behavior under failure}
To test the failure resilience of our system, we experimented with a faulty \dc that drops 2/3rds of all blocks it receives. This simulates a \dc with a lossy link. The time-series for the experiment is shown in \Cref{fig:dc_crash_test}.
In the beginning, we see that the throughput of the system faces no drops due to the lossy \dc, since the other two servers send the acknowledgments immediately.
Then we kill one healthy \dc number and bring it back up after some time.
During this phase, the throughput drops to almost half.
But it smoothly recovers when the healthy \dc comes back up.

% \begin{figure}
%     \centering 
%     \includegraphics[width=\linewidth]{figures/startup_fig.png}
%     \caption{Caption}
%     \label{fig:startup}
% \end{figure}

% \textbf{Insight: } 

% \subsection{End-to-End}

% \textbf{Insight: } 
\subsection{Case Study: Distributed and Privacy Preserving Deep Neural Network Training}

We demonstrate the applicability and scalability by training a Deep Neural Network with  multiple workers. Workers are connected by \algname with a shared memory paradigm.  and how scaling multiple numbers of workers can finish computing long, intensive tasks, we see how PSL can scale to do secure deep neural network training. 

We train a language model on Eurlex~\cite{chalkidis-etal-2019-large}, that has more than 1 million parameters with ADAM optimizer~\cite{zhang2018improved}. We port the Sub-linear Deep Learning Engine (SLIDE)\cite{DBLP:journals/corr/abs-1903-03129}, a CPU-based deep learning algorithm that utilizes multi-threading to reduce training time. The original SLIDE reports better performance than training with GPUs. We implement the data loader that fetches the training set and the SLIDE code from the \dc. The parameters of the deep learning model, such as biases and weights, are exchanged through  a shared memory array abstraction. 
The per-epoch latency is collected for 10 epochs.  We also compare these benchmarks to compiling SLIDE with -O3 and running natively. 

\begin{figure}
    \centering
    \includegraphics[width=\linewidth]{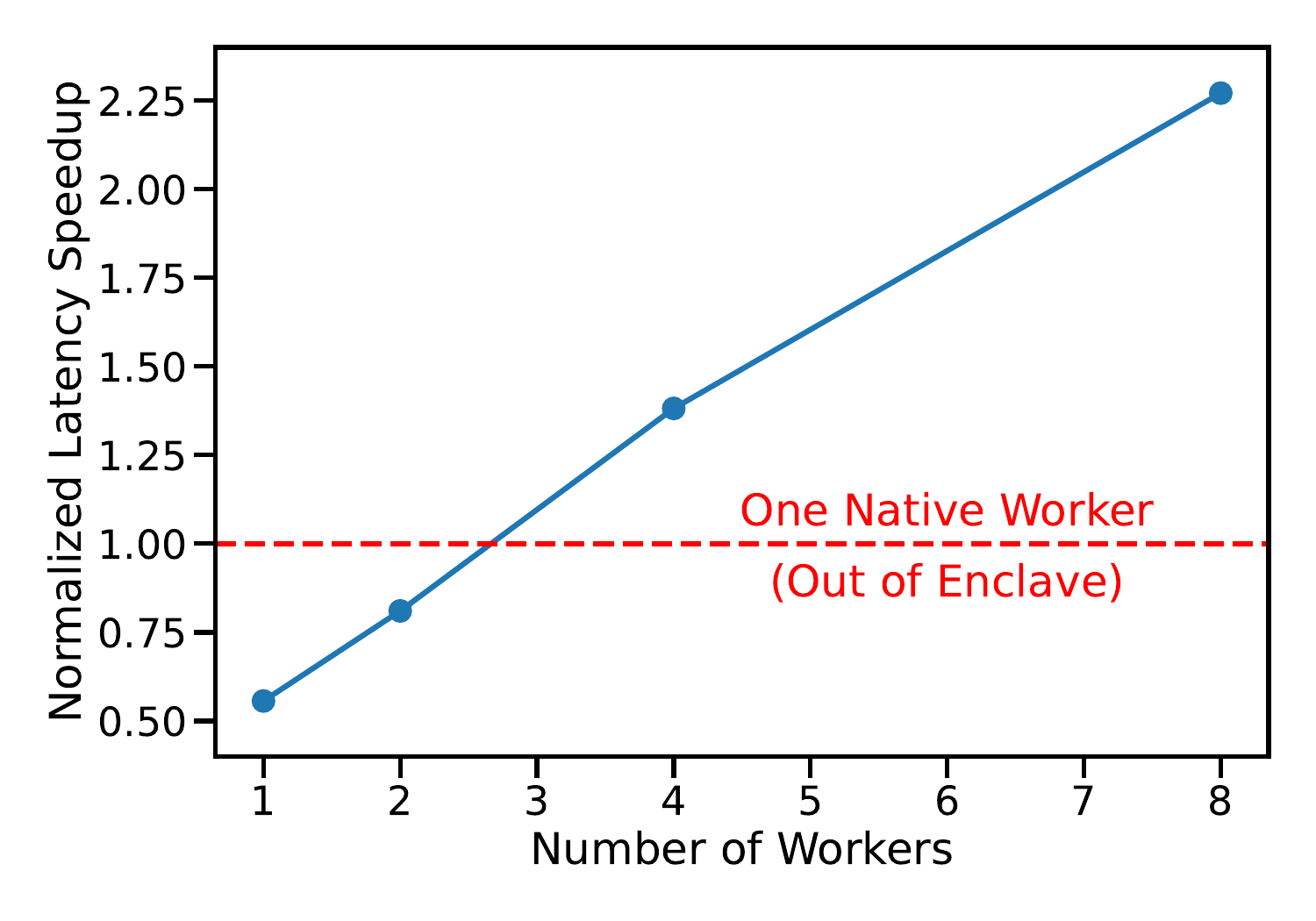}
    \vspace*{-0.35in}
    \caption{\textbf{Normalized Latency Speedup of Distributed Deep Learning Training with increasing number of workers.} \algname is only two times slower compared to a native worker with no enclave protection and WASM sandboxing. The speedup is linear with increasing number of workers in this domain.   }
    \label{fig:eval:dl}
    \vspace*{-0.1in}
\end{figure}

Figure \ref{fig:eval:dl} shows that \algname  achieves almost linear scaling as we scale from 1, 2, 4, and 8 worker nodes on distributed training latency. Furthermore, 
\algname only introduces 2x overhead compared against running with an unsecured and unsandboxed native worker, despite (1) the overhead of publishing the training data and weights to the \dc, (2) the overhead of SGX in data confidentiality, integrity and isolation, and (3)  the overhead of WASM sandbox, which provides dynamic safety checks\cite{234914}. %, more instructions, and more memory operations . 

% LLM is important, and security is important to abide certain policies. Enclave is the solution, but existing single-enclave cannot do it due to the EPC limit. 

% \textbf{Task} We train a model with XXX parameters, with XXX network architecture with Adam. We use X number of workers. 

\section{Related Work}
\label{sec:relworks}
% \eric{related work from psl paper}
\paragraph{Current Frameworks for FaaS}
There are existing stateless cloud-based FaaS implementations that support different forms of isolation. There is container based isolation like OpenWhisk~\cite{djemame2020open} and OpenFaaS \cite{OpenFaaS} that are stateless. There is also more lightweight forms of isolation like AWS Lambda \cite{lambda} which is based on the Firecracker MicroVM \cite{246288}. Faasm \cite{shillaker2020faasm} is the most similar to us as they use a WASM runtime engine, but do not run inside secure hardware. An earlier implementation, Secure Concurrency Layer~\cite{chen2022scl}, that exchanges states on a peer-to-peer network~\cite{chen2023fogros2}, only works for Intel SGX with limited programming language support.

% Existing cloud-based FaaS implementations, such as AWS Lambda \cite{lambda}, OpenWhisk~\cite{djemame2020open}, OpenFaaS \cite{OpenFaaS}, underutilize computing resources on the edge of the network. 
% Attempts to deploy such frameworks to the edge, such as Akamai \cite{akamai}, do not deliver the security guarantee required by the Edge Computing. S-FaaS \cite{alder2019s}, Clemmys \cite{trach2019clemmys} uses TEE  and cryptographic attestation to protect the confidentiality of the execution.  . %Naive extensions on these 

\paragraph{Secure Execution with TEE} 
There is existing FaaS based secure hardware that focus on optimizing specific metrics like cold start latencies by reusing enclave \cite{10.5555/3620237.3620462} or enable scalable memory sharing to securely share FaaS runtimes \cite{fengscalable, 10.1145/3548606.3560595, yu2020elasticlave}. In contrast, PSL attempts to build an entire FaaS framework with stateful computation. S-FaaS \cite{alder2019s} and Occulum \cite{shen2020occlum} proposes a FaaS framework on top of secure hardware. S-FaaS introduces transitive attestation similar to PSL and contributes a secure resource measurement to bill clients accurately using Intel TSX extensions. Occulum uses SFI to create light-weight enclaves, but only support stateless computation and also rely on specific Intel hardware extensions.

\paragraph{Consistency protocols on TEE}
Consistency protocols on TEEs generally fall into two categories:
(1) running the whole protocol inside enclave, e.g., CCF\cite{ccf_new_vldb},
(2) using enclaves as trusted endorsers, e.g., Nimble\cite{nimble}.
Although projects in the first category have larger TCB sizes, the confidentiality guarantees are also higher. Hence, \algname uses a similar approach.
A similar line of work tries to adapt known consensus protocols into TEEs. \cite{engraft, minbft}.
TEEs provide the non-equivocation guarantee which helps CFT protocols give BFT guarantees.
However, naively using this leads to easy responsiveness attacks. FlexiTrust protocols \cite{flexi} solve this issue by using more nodes but provide better performance.
However, all these works guarantee total order at the cost of parallelism.
To the best of our knowledge, \algname is the first to formally consider performant eventual consistency in a TEE-based system with asynchronous network.

\section{Conclusion and Future Work}
\label{sec:fut_work}
In this work, we present \algname, a lightweight, secure and stateful FaaS framework in TEE that supports WASM with JIT compilation which scales to multiple workers a provide 95k ops/s write throughput. It ensures secure state persistence and scalable and efficient eventual state consistency. We show a case study of distributed training on confidential data with PSL to show its adaptivity and scalability.  We leave it as future work to verify JIT transformations with a formal security guarantee.
We leave addressing rollback attacks, provenance tracking, and long-term key storage as future works.

We leave it as future work to reduce cold start latency of JIT by WASM runtime swapping: upon a cold start, the code starts an interpreted version of WAMR, while the JIT engine compiles code in the background.
In addition, we assume JIT engine is trusted in transforming WASM to machine code. 
To fully trust the transformations, one can create a JIT engine, such as \cite{10.1145/3385412.3385964} on creating verified binary lifters, that only transforms verified transformations in the code. 
In future work, we would also like to compare our performance against state-of-the-art TEE-based consensus protocols, such as MinBFT\cite{minbft}, FlexiBFT\cite{flexi}.
% We acknowledge the following limitations: 
% \begin{enumerate}
%     \item liveness - a property by design. We treat as future work 
%     \item since communication is broadcast, it becomes a logical next step to do hardware multicast 
% \end{enumerate}

% \input{chapters/related_work}
% \input{chapters/design}
% \input{chapters/future_work}
% \bibliographystyle{IEEEtran}
% \bibliography{IEEEabrv,references}

% \renewcommand*{\bibfont}{\footnotesize}
% \renewcommand{\baselinestretch}{0.9} %only do this for refs if crunched for space
\bibliographystyle{ACM-Reference-Format}
\bibliography{references}

\end{document}